\documentclass{emulateapj}
\usepackage{graphicx}

\shorttitle{Evershed flows and convection in penumbrae}
\shortauthors{Scharmer et al.}

\begin{document}

\title{Convection and the origin of Evershed flows in sunspot penumbrae}

\author{G.B Scharmer}
\affil{Institute for Solar Physics, Royal Swedish Academy of Sciences,
  AlbaNova University Center, SE-106\,91 Stockholm, Sweden}
\author{\AA. Nordlund}
\affil{Niels Bohr Institute, University of Copenhagen, Juliane
Maries
  Vej 30, 2100 Copenhagen, Denmark}
\and
\author{T. Heinemann}
\affil{DAMTP,
  Centre for Mathematical Sciences, Wilberforce Road, Cambridge CB3
  0WA, United Kingdom}

\begin{abstract}
  We discuss
  a numerical 3D radiation-MHD simulation of penumbral fine
  structure in a small sunspot. This simulation
  shows the development of short filamentary structures with horizontal
  flows, similar to observed Evershed flows, and an inward propagation
  of these structures at a speed compatible with observations.
  Although the lengths of these filaments are much shorter than observed,
  we conjecture that this simulation qualitatively reproduces the mechanisms
  responsible for filament formation and Evershed flows in penumbrae.
  We conclude that the Evershed flow represents
  the horizontal flow component of overturning convection in gaps with
  strongly reduced field strength. The top of the flow is always directed
  outward---away from the umbra---
  because of the broken symmetry due to the inclined magnetic field.
  Upflows occur in the inner parts of the gaps and
  most of the gas turns over radially (outwards and sideways), and descends
  back down again.  The ascending, cooling
  and overturning flow tends to bend magnetic field lines down,
  forcing a weakening of the field that makes it easier for gas located
  in an adjacent layer---further in---to initiate a similar sequence of
  motion, aided by lateral heating, thus causing the inward propagation of the filament.

\end{abstract}

\keywords{sunspots -- magnetic field}

\section{Introduction}

Only a convective process close to the
visible surface can explain the substantial radiative heat flux of sunspot penumbrae.
But spectroscopic studies, starting with Evershed (1909), show that penumbral
flows are predominantly horizontal and increase in
strength toward the outer penumbra. This gradually led to the view that Evershed
flows have an origin that is unrelated to convective processes beneath the
surface.

Some of the models that have been proposed
to explain Evershed flows rely on the concept of
nearly horizontal flux tubes, embedded in a more vertical magnetic field, and
with a (siphon like) flow driven along the flux tube by enhanced gas pressure
at one end of its ends (e.g.\ Schlichenmaier 2002).
Such ``siphon flow'' models were
first proposed by Meyer \& Schmidt (1968) and were further developed by e.g.
Thomas (1988)
and by Thomas and Montesinos (1991). Schlichenmaier et al. (1998a,b) and
Schlichenmaier (2002), using 1D MHD
simulations, modeled time dependent flows in thin flux tubes
embedded in an atmosphere of given properties.

However, nearly horizontal flux tubes cannot carry the heat flux
needed to explain the radiative losses of the penumbra (Schlichenmaier \& Solanki 2003).

A different view of penumbra fine structure emerged from the discovery that
bright penumbra filaments have dark cores (Scharmer et al. 2002). These dark
cores were proposed by Spruit and Scharmer (2006) to be a consequence of
nearly field-free gaps below the visible surface. The darkness of the cores is
explained as the result of locally enhanced gas pressure associated with the
gaps combined with an overall drop of temperature with height.
Within the gaps, overturning convection transports energy to
the surface, providing an explanation for the penumbra heat flux. This "gappy"
model explains the strongly inclined field at the locations of the dark cores
as a natural consequence of a near-potential magnetic field folding over the gap.
Magnetostatic models of this type are able to explain observed differences
between the inner and outer penumbra (Scharmer \& Spruit 2006).

Recently, Heinemann et al. (2007) presented the first 3D MHD simulations
relevant to the formation of penumbral fine structure. The simulations show the
formation of short filamentary structures. These filaments reproduce morphological,
dynamic and evolutionary properties that agree reasonably well with observations.
We conjecture that the simulations capture the essential physics of such filaments,
even though their lengths are much shorter than observed.

In the current Letter we
use these simulations to make inferences about the nature of convection and
Evershed flows in sunspot penumbrae. One of the main conclusions is that
the Evershed flow represents the horizontal flow component of overturning
convection in penumbrae.

\section{MHD Simulations}

The simulations of Heinemann et al. (2007) were carried out using the PENCIL
code\footnote{see
http://www.nordita.org/software/pencil-code}, modified to handle energy
transfer by radiation in a grey atmosphere (Heinemann 2006; Heinemann
et al.\@ 2006). We used a rectangular computational box
$12448\times6212$~km in the horizontal ($x$ and $y$)
directions, extending over a depth ($z$) range of $3094$~km and
with a grid separation of $24.36$~km in both horizontal and vertical
directions ($512\times256\times128$ grid points). The quiet sun
photosphere is located approximately $700$~km below the upper
boundary.

We emphasize the following
agreement between the simulations and observed properties of penumbra filaments:
1) The association of dark cores with locally reduced field strength and
   a more inclined magnetic field.
2) The presence of outward horizontal flows in a thin layer close to the
   visible surface.
3) The presence of strong upflows at the innermost point of the filamentary
   structures.
4) The inward migration of these structures toward the umbra.

\subsection{Relation to umbral convection simulation}

The penumbra simulations show similarities with the umbral convection
simulations of Sch\"ussler and V\"ogler (2006). Their simulations demonstrate the
development of narrow upflow plumes which become nearly field-free near the
surface layers, and horizontal flows cospatial with dark lanes, similar to
the dark cores seen in our simulations. Convective downflows are concentrated
at the endpoints of the dark lanes, which sometimes split in Y-shapes. Notably, the gaps
in the umbra simulations do not extend to the bottom of the simulation box but are
limited to a depth of about 500~km below the surface. The gaps formed in
the penumbra simulations are similar, but aligned with an {\em inclined} field.
The associated horizontal flows are always in the direction of the surrounding field-free
photosphere.

The Y-shaped dark lanes seen in umbral dots in Sch\"ussler and V\"ogler's simulations
are similar to observed peripheral umbral dots connected to dark-cored penumbral
filaments (Langhans et al. 2007). Recently, dark lanes in umbral fine structure have
also been reported by Bharti et al. (2007) and Rimmele (2008). There is thus already
observational support for these simulations, but the flows along dark lanes and downflows
at their ends in the umbral dot simulations have not yet been observed.

\subsection{Relation to light bridge simulations}

Simulations of light bridges by Nordlund (2006) and Heinemann (2006)
share some crucial properties with the umbral dot simulations.  The brightness is
supported by convective heat transport, which here is able to push open and maintain
a ``gap'' with strongly reduced magnetic field strength. The cusp shaped
magnetic field arching over the gap, which extends for a considerable distance,
is associated with a dark core along its symmetry line. Such light bridge dark
cores that continue as penumbra dark cores have been observed (Scharmer et al. 2007),
suggesting a common origin.

\section{Penumbra convection and Evershed flows}

\begin{figure}[htbp]
\includegraphics[width=\columnwidth]{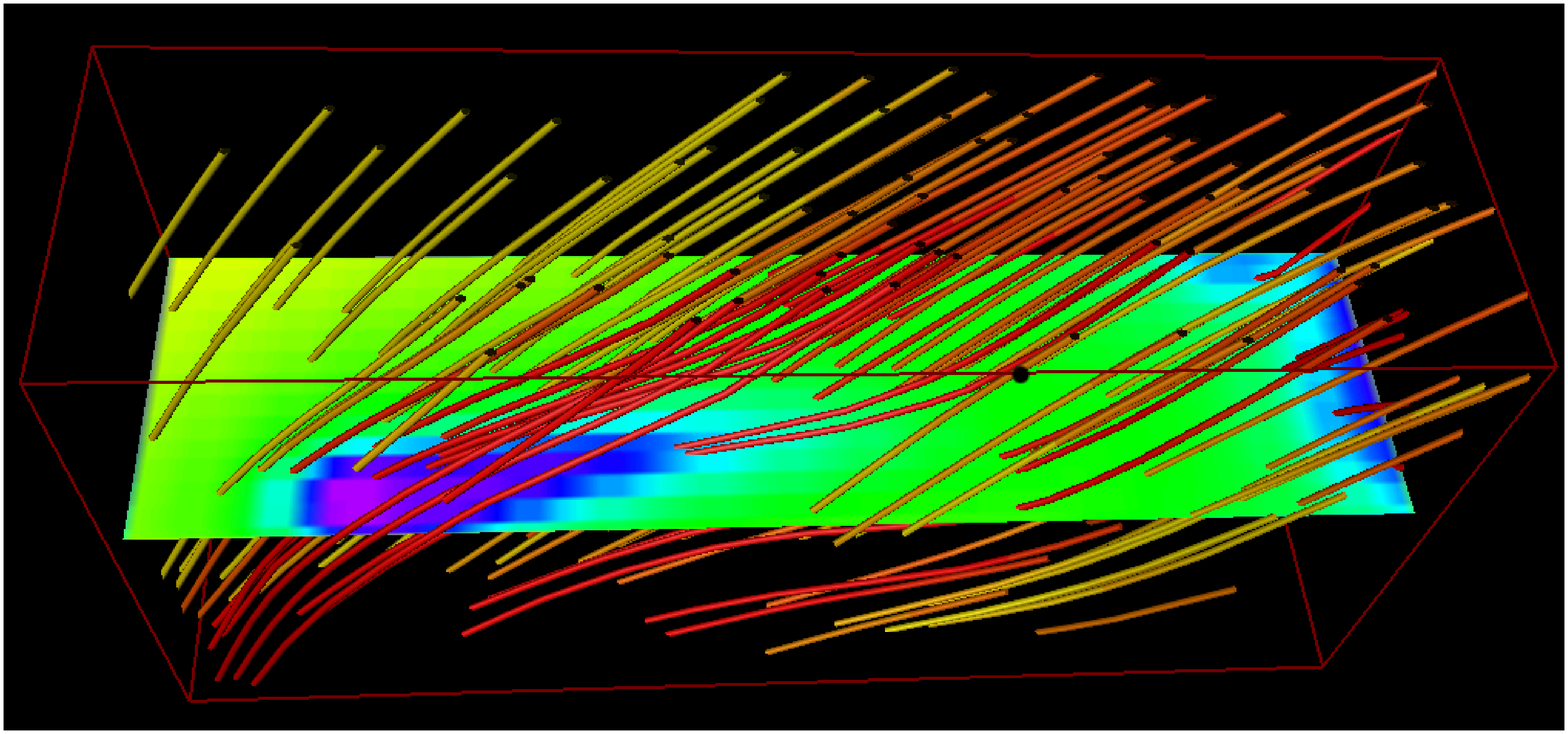}\\[1pt]
\includegraphics[width=\columnwidth]{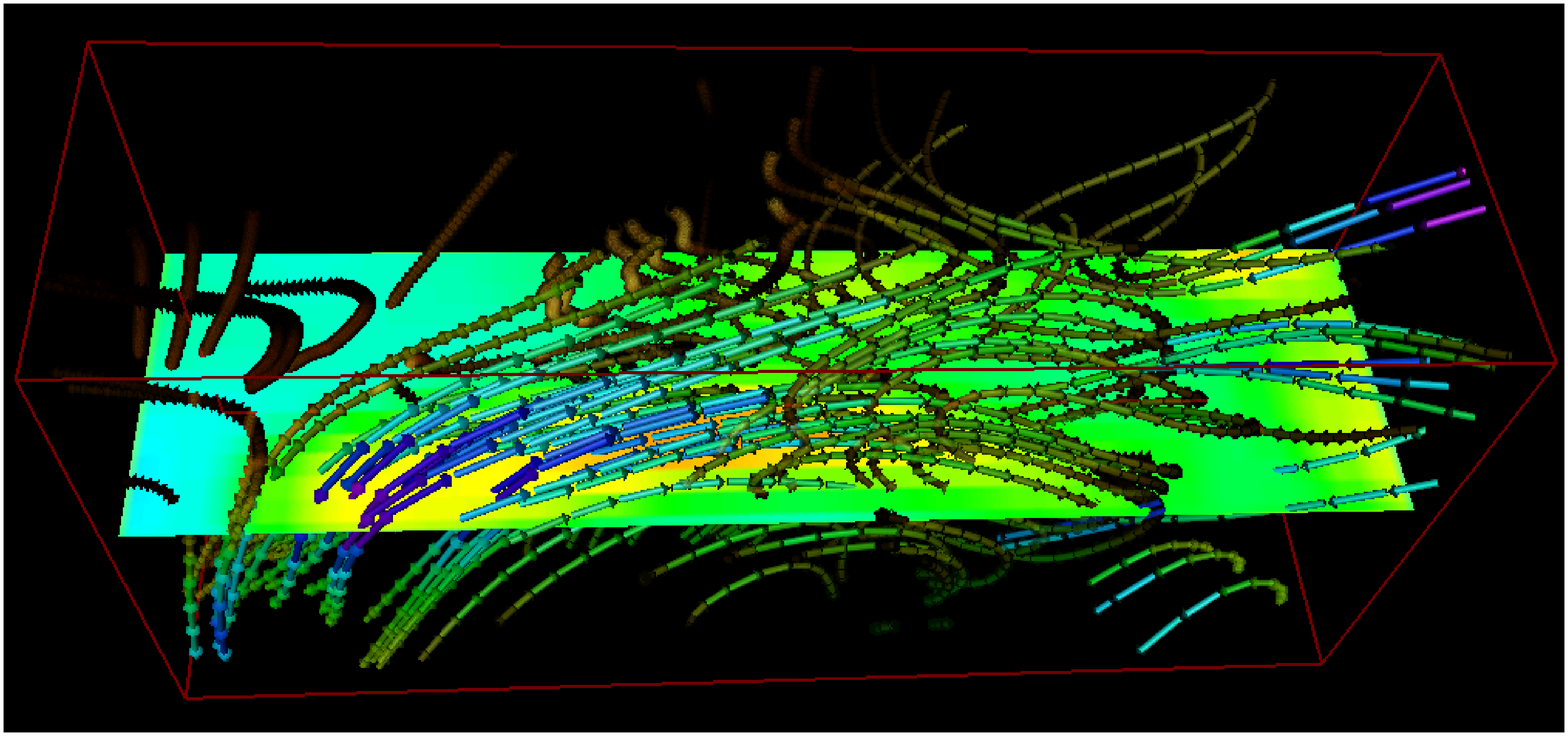}
  \caption{\small
  The upper panel shows the vector magnetic field, color coded with magnetic
  field strength, with a cutting plane containing a color coded image of temperature.
  Purple to blue represents large values, orange to red represent small values.
  The lower panel shows the vector velocity field, color coded with velocity
  magnitude, and with a cutting plane color coded image showing the strength
  of the vertical magnetic field component.
  The size of the box is $\sim 1200\times 340\times 550$ km.
  \\}
  \label{gapflow}
\end{figure}

From field-free (photospheric) convection simulations (e.g. Stein \& Nordlund
1998) we can estimate that hot gas needs to reach the visible surface
with a vertical speed of the order of 1--2 km/s
in order to support and maintain the observed average luminosity
of penumbrae. Strong such upflows have not yet been observed in penumbrae, except
at the innermost parts of filaments (Rimmele and Marino 2006). The simulations
demonstrate, as expected, that upflows occur in the inner parts of the gaps and downflows
at their sides. The small azimuthal separations between the upflows and downflows
would make such flows difficult to observe, except at their innermost parts
where there are no downflows.

In the quiet sun, convection is associated with horizontal flows that
are typically twice as strong as the vertical
flows and occasionally reach supersonic velocities near the
photosphere (Stein and Nordlund 1998).  The analogous structure of the
convective flows that we observe in the penumbra simulations prompt us to {\em identify the
observed Evershed flows with the horizontal flow component of overturning convection
in penumbrae}. Their visibility is then clearly due to their unidirectional nature---they are
always directed away from the umbra.
In the simulations, the horizontal outward flows in the optically visible
parts of the gaps increase towards $\tau=1$, which is also where the
efficiency of radiative cooling peaks.
It seems well established that the Evershed flow speed increases with optical depth (e.g.,
Bellot Rubio et al. 2006), so the simulation results are consistent with observations also
in this respect.

\subsection{Convective flows and magnetic field interaction}

Figure 1 shows a detail of a snapshot from the penumbra simulation of
Heinemann et al. (2007).  The flow field and the magnetic field are
visualized in a small box centered on one of the inward
propagating penumbral filaments (the umbra is situated
to the left of the box).

In the foreground one can see that the magnetic field is bent
down, resulting in a locally increased inclination, and also in a
local weakening of the field strength (encoded as more reddish
color).  The cause of this effect can be appreciated by
considering the shape of the velocity field, which is illustrated
in the lower panel of the figure.  As shown by the purple color,
the strongest velocity field occurs to the left, at the inward
propagating head of the filament.  As required by conservation of
mass the ascending flow forces an overturning motion, which pushes
the magnetic field aside and forces field lines to bend over
outwards (the cutting plane shows yellow / orange weaker vertical
field).

Since these flows occur on a scale not much larger than the numerical
resolution of the simulation the magnetic diffusivity is significant,
and the magnetic field lines are thus able to partly slip
through the flow, back towards their initial arrangement.  Nevertheless,
the effect of the flow on the magnetic field lines is obvious enough,
and would only become stronger with decreasing magnetic diffusivity.  In
the limit of very small magnetic diffusivity one would expect that the
flow would force open a nearly field-free narrow channel, with overturning
convection inside.  The flow pattern within such a channel would be quite
similar to the flow pattern illustrated in Fig.\ 1; the velocity field
must be similar in shape and amplitude to be able to transport the
necessary amount of energy up to the visible surface.

Given the necessity of horizontal flows inside the channels, and
given the necessity to expose the hot gas to the surface cooling for
a sufficiently long time for it to loose buoyancy and start to descend,
it is clear that flows that extend {\em along} the channels (and
hence also more or less along the magnetic field) are
energetically as well as topologically favorable.
So, while there is also a flow component {\em across}
filaments, away from the center of the filament and with cooler
downflows occurring near the boundaries of the gaps, the main
flow direction must be {\em along} the filament.

Another way of seeing this is to consider the pressure pattern
associated with the flow.  The horizontal flow is primarily
driven by the pressure difference between the upflow and downflow
points.  That pressure difference is, in turn, a consequence of
the cooling that occurs as the gas reaches the surface and flows
between the points of ascent and descent.

An important consequence of the surface cooling is that the gas becomes
denser and heavier, and that it therefore more efficiently bends the
field lines downward. This creates near-horizontal field lines that
in turn can more easily accommodate horizontal flows.  That tendency
is clearly seen already in the current simulation, and must be even
stronger in the real penumbra.  Most likely this effect
contributes to the ``uncombed'' structure of the penumbral magnetic
field, with some field lines being much more inclined than would be
expected from a near-potential situation.

The lower numerical resolution in these penumbra simulations--- $\sim$24
km in all directions, as compared to 10--20 km (vertical and
horizontal) in the umbral simulations of Sch{\"u}ssler and V{\"o}gler
(2006)---probably explains why the magnetic field strengths in the gaps
are relatively higher in the penumbra simulations than in the umbra simulations.

Note that the amplitudes of vertical and horizontal flows are to a large
extent determined by energy and mass flux requirements and must be robust
properties of the flow; improvements in numerical resolution are expected to
primarily cause changes in the magnetic field, which is then forced to
locally become more perfectly aligned with the flow.  However, lower
diffusivity also leads to the development of more small scale structure
and a more chaotic behavior of the magnetic field lines, so it is likely
that some level of ``turbulent diffusivity'' will always be present.

\subsection{The structure and migration along the gaps}

An important question concerns the inward migration of the gaps.
The field lines that are bent down by the ascending and
cooling gas cause a reduction of the
field strength in the upper parts of field lines that
connect to the {\em inward} side of the gap, making it easier
for gas to come up to the surface on that side. Gaps pushed open by the
process described above therefore have a tendency to propagate in
the opposite direction of the horizontal flow itself.

This
tendency is encouraged additionally by the anisotropy of heat
transport near the ``head'' of the gap, in the following sense:
In order to rise into the "opening" the gas must be heated from
the umbral-like temperature towards the photospheric one.  That
heating is mainly due to radiation that ``leaks out''
sideways, from the edge of the convectively heated gas inside
the gap into the less dense and more transparent gas immediately
outside.

The magnitude of the horizontal heat flux
that is available is a fraction of the nominal
vertical solar flux.   That amount of flux can support a certain
rate of heating per unit volume and time, corresponding to a
rather well defined propagation speed that does not too much
on other details.   But the propagation becomes
increasingly difficult with stronger and more vertical fields, so
the inward migration of the gap must eventually come to a stop.
Beyond that, there is a transition to the umbral dot behavior,
defining thus the boundary between the penumbra and the umbra.

\section{Discussion}

Most present theoretical models of penumbra fine structure are based on the
concept of embedded flux tubes.
For nearly a decade, the moving tube models (Schlichenmaier et al. 1998a,b;
Schlichenmaier 2002) and schematic embedded flux tube models (Solanki and Montavon 1993)
have been seen to mutually support each other.
We have argued that this
congruence is due to a combination of oversimplified models and non-unique
interpretations of polarized spectra and images (Spruit \& Scharmer 2006, Scharmer \&
Spruit 2006).

The limitations of these models are indeed fundamental:
Attempts to construct models of magnetostatic flux tubes
with given (round) cross sections, embedded in a surrounding more vertical field,
have been made by Borrero (2007). This leads to an overconstrained problem
where magnetostatic equilibrium determines not only the gas pressure, but also the density
and thus the temperature within the flux tube. There is no room for an energy
equation in these models---the temperature distribution is already determined by the
assumed (round) cross section of the flux tube. Moreover, the temperatures actually
obtained in the models of Borrero (2007) are far too low---the predicted average penumbra
intensity is approximately a factor three lower than observed. This highlights the
difficulties of satisfying requirements of near-magnetostatic equilibrium with siphon flow
and moving tube models; they
also generally fail to transport enough heat to the surface.

Even ignoring these difficulties, the agreement between observed and
calculated net circular polarization that can be obtained with such models (Borrero et al. 2007a) does not
specifically support the {\em existence} of penumbra flux tubes, since the magnetic field of its
{\em observable} part is similar to that of a gap model (Scharmer \& Spruit 2006).
Also, the evidence recently found for field lines wrapping around bright
filamentary structures with strong flows and weaker and more horizontal
magnetic field (Borrero et al.\ 2007b) are consistent with both flux tube
and gappy models.

Recently, apparent motions of intensity structure
{\em across} penumbra filaments were observed for several sunspots away from disk center
(Ichimoto et al. 2007). These apparent motions were observed only in parts of the penumbra that
are perpendicular to the symmetry line and the motions were found to be always in the direction from
the limb toward disk center. These observations are difficult to reconcile with (twisting)
flux tubes but are consistent with overturning convective flows in gaps (Spruit \& Scharmer
2006), with the limb side of the gap hidden by a strongly warped $\tau=1$ surface
(Scharmer \& Spruit 2006). Such convective flow patterns are indeed seen in the
simulations discussed here.

\section{Conclusions}

In this Letter we present a scenario where the Evershed flow is not only related to,
but actually is {\em identical} to the horizontal component of  overturning convection in
penumbrae.  The observed migration of filament heads towards the umbra is reproduced by
the numerical model, and is found to be the
result of a pattern motion, with an upflow on the {\em inward} side enabled by
the bending down of field lines by gas that is cooled at the surface and therefore becomes
heavy. This process is aided by lateral radiative heating.

One of the main strengths of this model is that the energy transport and energy
balance is at center stage, rather than being an embarrassment as in some other
models. In the present scenario energy transport
is the controlling factor, relegating issues such as magnetic diffusion
to only secondary roles.  It is the strong cooling of gas at the surface (seen
observationally as the relatively large luminosity of the penumbra) that provides
the mechanical driving of the flows, via the resulting pressure differences and
pressure gradients.  The flows in turn force the magnetic field lines apart and
open up the channels that are seen as dark-cored filament in high-resolution
observations.  In that context the exaggerated magnetic diffusivity in the numerical
models mainly has the effect to allow the field lines to slip back through the
plasma somewhat too easily; this doesn't change the overall picture, which is made
robust precisely by the strong constraints from the energy transport requirements.

In addition to providing a satisfactory explanation of penumbra fine structure, the
present interpretation also allows a unified view of convection and magnetic field
interaction in penumbrae (Spruit \& Scharmer 2006, Heinemann et al.\ 2007), umbrae
(Sch\"ussler and V\"ogler 2006) and light bridges (Heinemann 2006).
Whereas better observations and simulations may be needed to firmly establish this view,
the present 3D MHD simulation already provides a fundamentally more consistent
representation of penumbra dynamics and filament formation than
1D flux tube and two-component models.

\begin{acknowledgments}
TH thanks Nordita and DAMTP for financial support.
{\AA N} acknowledges support from the Danish Natural Science Research Council (FNU).
Computing time provided by the Danish Center for Scientific Computing (DCSC) is
gratefully acknowledged.  The figure panels were produced with VAPOR (www.vapor.ucar.edu).
\end{acknowledgments}

\end{document}